\documentclass{PoS}

\newcommand{\nc}[1]{\newcommand{#1}}

\nc{\its}[1]{\itshape #1 \upshape}
\nc{\mc}[3]{\multicolumn{#1}{#2}{#3}}

\nc{\bc}{\begin{center}}
\nc{\ec}{\end{center}}
\nc{\ig}[1]{\bc \includegraphics{#1} \ec}

\nc{\ben}{\begin{enumerate}}
\nc{\een}{\end{enumerate}}

\nc{\bo}[1]{\mbox{\boldmath \( #1 \! \! \)  \unboldmath}}

\nc{\be}{\begin{eqnarray}}
\nc{\ee}{\end{eqnarray}}
\nc{\bew}{\begin{eqnarray*}}
\nc{\eew}{\end{eqnarray*}}

\nc{\nnn}{\nonumber}
\nc{\f}[2]{\frac{#1}{#2}}
\nc{\td}[2]{\f{d #1}{d #2}}
\nc{\pd}[2]{\f{\partial #1}{\partial #2}}
\nc{\suli}{\sum\limits}
\nc{\proli}{\prod\limits}
\nc{\ili}{\int\limits}
\nc{\sr}[2]{\stackrel{#1}{#2}}
\nc{\dps}{\displaystyle}
\nc{\ket}[1]{\left| #1 \right>}
\nc{\bra}[1]{\left< #1 \right|}
\nc{\bracket}[2]{\left< #1 \right| \left. \! #2 \right>}
\nc{\norm}[1]{\left\| #1 \right\|}
\nc{\lndm}[1]{\pd{^{#1} \ln{\det{D}}}{\mu^{#1}}}
\nc{\pdmm}[1]{D^{-1} \pd{^{#1} D}{\mu^{#1}}}
\nc{\pdm}{D^{-1}\pd{D}{\mu}}
\nc{\trac}[1]{\mbox{Tr}\left(#1\right)}

\nc{\la}{\langle}
\nc{\ra}{\rangle}
\def\simge{\mathrel{%
       \rlap{\raise 0.511ex \hbox{$>$}}{\lower 0.511ex \hbox{$\sim$}}}}
\def\simle{\mathrel{
       \rlap{\raise 0.511ex \hbox{$<$}}{\lower 0.511ex \hbox{$\sim$}}}}
\nc{\Vsi}{V_{\bf 1}}
\nc{\Vav}{V_{\bf av}}
\nc{\tr}{{\rm Tr}}
\nc{\bx}{{\bf x}}
\nc{\by}{{\bf y}}
\nc{\bz}{{\bf 0}}
\nc{\Te}{{\cal T}}
\nc{\C}{C}
\nc{\Tpc}{T_{\rm pc}}
\nc{\mpmv}{m_{\rm PS}/m_{\rm V}}
\nc{\Ooe}{\Omega_{\rm oe}}
\nc{\Ooo}{\Omega_{\rm oo}}
\nc{\Cee}{C^{\rm ee}}
\nc{\Ceo}{C^{\rm eo}}
\nc{\Coe}{C^{\rm oe}}
\nc{\Coo}{C^{\rm oo}}
\nc{\Ceeoo}{C^{\rm e/o}}
\nc{\meo}{m_{\rm e/o}}
\nc{\dg}{\dagger}
\nc{\Qb}{\bar{Q}}
\nc{\mq}{\mu_{q}}
\nc{\Om}{\Omega_{\rm M}}
\nc{\Oe}{\Omega_{\rm E}}
\nc{\Ome}{\Omega_{{\rm M},+}}
\nc{\Omo}{\Omega_{{\rm M},-}}
\nc{\Oee}{\Omega_{{\rm E},+}}
\nc{\Oeo}{\Omega_{{\rm E},-}}

\nc{\Cmef}{C_{{\rm M},+}^{\rm F}}
\nc{\Cmof}{C_{{\rm M},-}^{\rm F}}
\nc{\Ceef}{C_{{\rm E},+}^{\rm F}}
\nc{\Ceof}{C_{{\rm E},-}^{\rm F}}
\nc{\Cmei}{C_{{\rm M},+}^{\rm I}}
\nc{\Ceoi}{C_{{\rm E},-}^{\rm I}}

\nc{\Cmf}{C_{{\rm M}}^{\rm F}}
\nc{\Cef}{C_{{\rm E}}^{\rm F}}
\nc{\Cmi}{C_{{\rm M}}^{\rm I}}
\nc{\Cei}{C_{{\rm E}}^{\rm I}}

\nc{\Mmef}{m_{{\rm M},+}^{\rm F}}
\nc{\Mmof}{m_{{\rm M},-}^{\rm F}}
\nc{\Meef}{m_{{\rm E},+}^{\rm F}}
\nc{\Meof}{m_{{\rm E},-}^{\rm F}}
\nc{\Mmei}{m_{{\rm M},+}^{\rm I}}
\nc{\Meoi}{m_{{\rm E},-}^{\rm I}}

\title{Magnetic and electric screening masses from Polyakov-line correlations }

\ShortTitle{Magnetic and electric screening masses from Polyakov-line correlations }

\author{\speaker{Yu Maezawa}$^a$,
S.~Aoki$^{b,c}$, S.~Ejiri$^d$, T.~Hatsuda$^e$, N.~Ishii$^f$, 
K.~Kanaya$^b$, N.~Ukita$^f$, and T.~Umeda$^b$ (WHOT-QCD Collaboration)
\\%
$^a$En'yo Radiation Laboratory, Nishina Accelerator Research Center,
    RIKEN, Wako, Saitama 351-0198, Japan
\\
$^b$Graduate School of Pure and Applied Sciences,
    University of Tsukuba, Tsukuba, Ibaraki 305-8571, Japan
\\
$^c$RIKEN BNL Research Center, Brookhaven National Laboratory, 
    Upton, NY 11973, USA
\\
$^d$Physics Department,
    Brookhaven National Laboratory, Upton, New York 11973, USA
\\
$^e$Department of Physics, The University of Tokyo, 
    Tokyo 113-0033, Japan
\\
$^f$Center for Computational Sciences,
    University of Tsukuba, Tsukuba, Ibaraki 305-8577, Japan
\\
        E-mail: \email{maezawa@ribf.riken.jp}}

\abstract{
Screening properties of the quark gluon plasma are
 studied from the Polyakov-line correlations 
 in lattice QCD simulations with two flavors of improved Wilson quarks 
 at temperatures $T/\Tpc \simeq 1$--$4$
 where $\Tpc$ is the pseudocritical temperature.
Using the Euclidean-time reflection symmetry
 and the charge conjugation symmetry, 
  we introduce various types of Polyakov-line correlations
  and extract screening masses in magnetic and electric sectors.
We find that a ratio of the screening masses in 
 the electric sector to the magnetic sector shows qualitative agreement
  with a prediction from the dimensionally-reduced effective field theory
   and the ${\cal N}=4$ supersymmetric Yang-Mills theory at $1.3 < T/\Tpc < 3$.}

\FullConference{The XXVI International Symposium on Lattice Field Theory \\
                 July 14 - 19, 2008\\
                 Williamsburg, Virginia, USA}

\begin{document}

\section{Introduction}

Electric and magnetic 
 screening masses of the gluon are the fundamental
 quantities of the quark-gluon plasma at high temperature 
\cite{Kraemmer:2003gd}.
 In this report,  we employ lattice QCD simulations with two flavors of 
  improved Wilson quarks at temperatures $T/\Tpc \simeq 1$--4
  and try to extract both screening masses
 from the correlations of the Polyakov-line operator
   classified under the 
  the Euclidean-time reflection 
 and the charge conjugation \cite{Arnold:1995bh}.
 Also, we make a comparison of our results with
  the predictions from the dimensionally-reduced effective field theory
   and the ${\cal N}=4$ supersymmetric Yang-Mills theory.

\section{Correlations of the Polyakov-line operator}

In order to extract the electric and magnetic screening masses,
 we  classify the Polyakov-line operator into different classes 
 on the basis of the symmetries under
  the Euclidean time-reflection \cite{Arnold:1995bh} ($\Te$) 
  and the charge conjugation ($C$).
 Both are good symmetries of QCD at zero chemical potential.
  (Note that $\Te$ corresponds to the product of 
 the time reversal $T$ and charge conjugation ${ C}$
 in the Minkowski space.)
 Under $\Te$, the gluon fields are transformed as,
\be
 \vec{A}(\tau,\bx) \rightarrow \vec{A}(-\tau,\bx) , \ \ 
 A_4 (\tau,\bx)    \rightarrow   - A_4(-\tau,\bx) .
\label{eq:TeA}
\ee
An operator belongs to 
 the magnetic (electric) sector if it is even (odd) under $\Te$. 
On the other hand,
 the charge conjugation changes the gluon field as,
\be
A_\mu(\tau,\bx) \rightarrow - A^*_\mu(\tau,\bx)
.
\ee
The Polyakov-line operator
\be
\Omega (\bx) = P \exp \left[ i g \int_0^{1/T} d\tau A_4(\tau,\bx) \right]
\ee
transforms under $\Te$ symmetry and $C$ symmetry as
\be
\Te : \ \ \Omega \rightarrow \Omega^\dag,
\quad
C : \ \   \Omega \rightarrow \Omega^\ast
.
\ee
Therefore, we define $\Te$-even (odd) operator by
 its Hermitian (anti-Hermitian) component:
\be
\Om \equiv \f{1}{2} (\Omega + \Omega^\dagger),
\quad
\Oe \equiv \f{1}{2} (\Omega - \Omega^\dagger),
\ee
where the subscript implies that the operator is in the magnetic (electric)
 sector. 
Furthermore, they can be decomposed into  $C$-even and $C$-odd operators as
\be
\label{eq:Omega-ME}
\Omega_{{\rm M},\pm} =
 \frac{1}{2} (\Om \pm \Om^* )
 , \ \ 
\Omega_{{\rm E},\pm} =
 \frac{1}{2} (\Oe \pm \Oe^* )
.
\ee

We can define two types of gauge invariant correlators
using Eq.(\ref{eq:Omega-ME}) as
\be
\Cmei (r,T) &\equiv&
 \langle \tr \Ome ({\bf x}) \tr \Ome ({\bf y}) \rangle
 -  \la \tr \Omega \ra^2
,\\
\Ceoi (r,T) &\equiv&
 \langle \tr \Oeo ({\bf x}) \tr \Oeo ({\bf y}) \rangle
,
\ee
where $r \equiv |\bx - \by|$, and the superscript ``I'' implies 
 that the correlators are gauge invariant.
Since  $\tr \Omo = \tr \Oee = 0$, the correlators of these 
operators are identically zero.
If we are allowed to fix gauge, 
we can construct four types of correlators:
\be
\Cmef (r,T) &\equiv& \langle \tr \Ome ({\bf x}) \Ome ({\bf y}) \rangle - \la \tr \Omega \ra^2, \\
\Cmof (r,T) &\equiv& \langle \tr \Omo ({\bf x}) \Omo ({\bf y}) \rangle, \\
\Ceef (r,T) &\equiv& \langle \tr \Oee ({\bf x}) \Oee ({\bf y}) \rangle, \\
\Ceof (r,T) &\equiv& \langle \tr \Oeo ({\bf x}) \Oeo ({\bf y}) \rangle.
\ee
where the superscript ``F'' implies gauge-fixed correlators.
In this study, we adopt the Coulomb gauge fixing.

 We define the screening masses of these correlators 
 through the fit with the following screened Coulomb form at long distances:
\be
C_{\Te,\C}^{\rm I(F)} (r,T) \rightarrow
 \alpha_{\Te,\C}^{\rm I(F)}(T) \f{e^{-m_{\Te,\C}^{\rm I(F)}(T) r}}{r}
\ee
where $\alpha_{\Te,\C}^{\rm I(F)}(T)$ and $m_{\Te,\C}^{\rm I(F)}(T)$
 are fitting parameters.
Then we can extract six screening masses all together,
\be
m_{\Te,\C}^{\rm I(F)} ({\rm Magnetic\ sector}) : && \Mmei,\ \Mmef,\ \Mmof
,\\
m_{\Te,\C}^{\rm I(F)} ({\rm Electric\ sector}) : &&  \Meoi,\ \Meef,\ \Meof
.
\ee

\section{Numerical simulations}
\label{sec3}

We employ a renormalization group improved gauge action and a clover improved
Wilson quark action with two flavors.
The simulations are performed on a lattice with a size of 
$N_s^3 \times N_t = 16^3 \times 4$
 along lines of constant physics, 
i.e. lines of constant $m_{\rm PS} / m_{\rm V}$
(the ratio of pseudoscalar and vector meson masses) at $T=0$
in the space of simulation parameters.
Details of the lines of constant physics with the same actions are
summarized in Refs.~\cite{cp1,cp2,Maezawa:2007fc}.
We take two values of $m_{\rm PS} / m_{\rm V}= 0.65$ and 0.80
with the temperature range of $T/ \Tpc \sim$ 1.0--4.0 (9 points)
and 1.0--3.0 (7 points), 
 respectively, where $\Tpc$ is the pseudocritical 
temperature along the line of constant physics.
The number of trajectories for each run after thermalization is 
5000--6000, and we measure physical quantities at every 10 trajectories.
The fits are performed by minimizing $\chi^2/N_{\rm DF}$ 
 with fit ranges of $0.35 \le rT \le 1.0$ for $\Ceoi$ and 
  $0.5 \le rT \le 1.25$ for other orrelators.

\begin{figure}[tbp]
  \vspace{-2mm}
  \begin{center}
    \begin{tabular}{cc}
    \includegraphics[width=74mm]{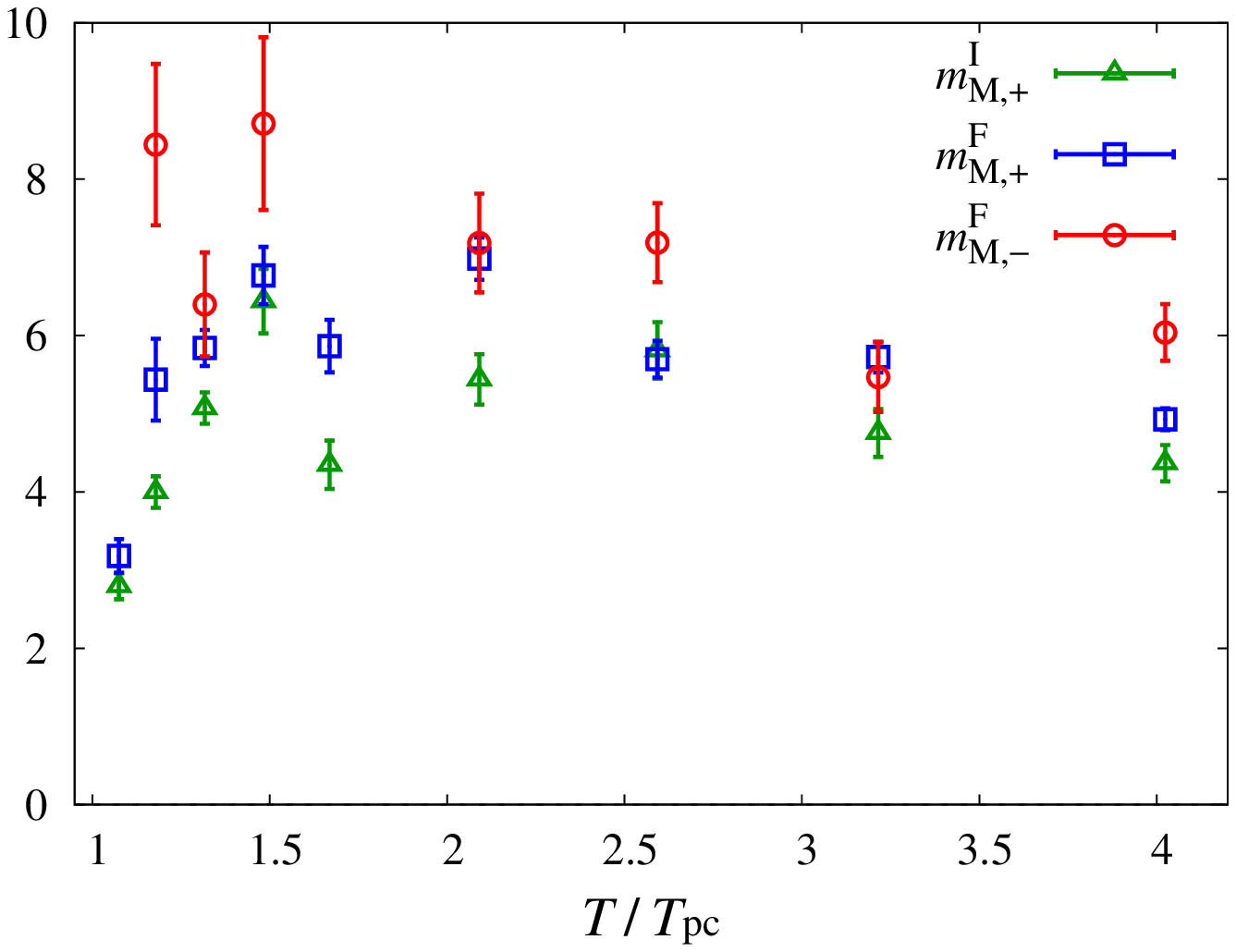} &
    \includegraphics[width=74mm]{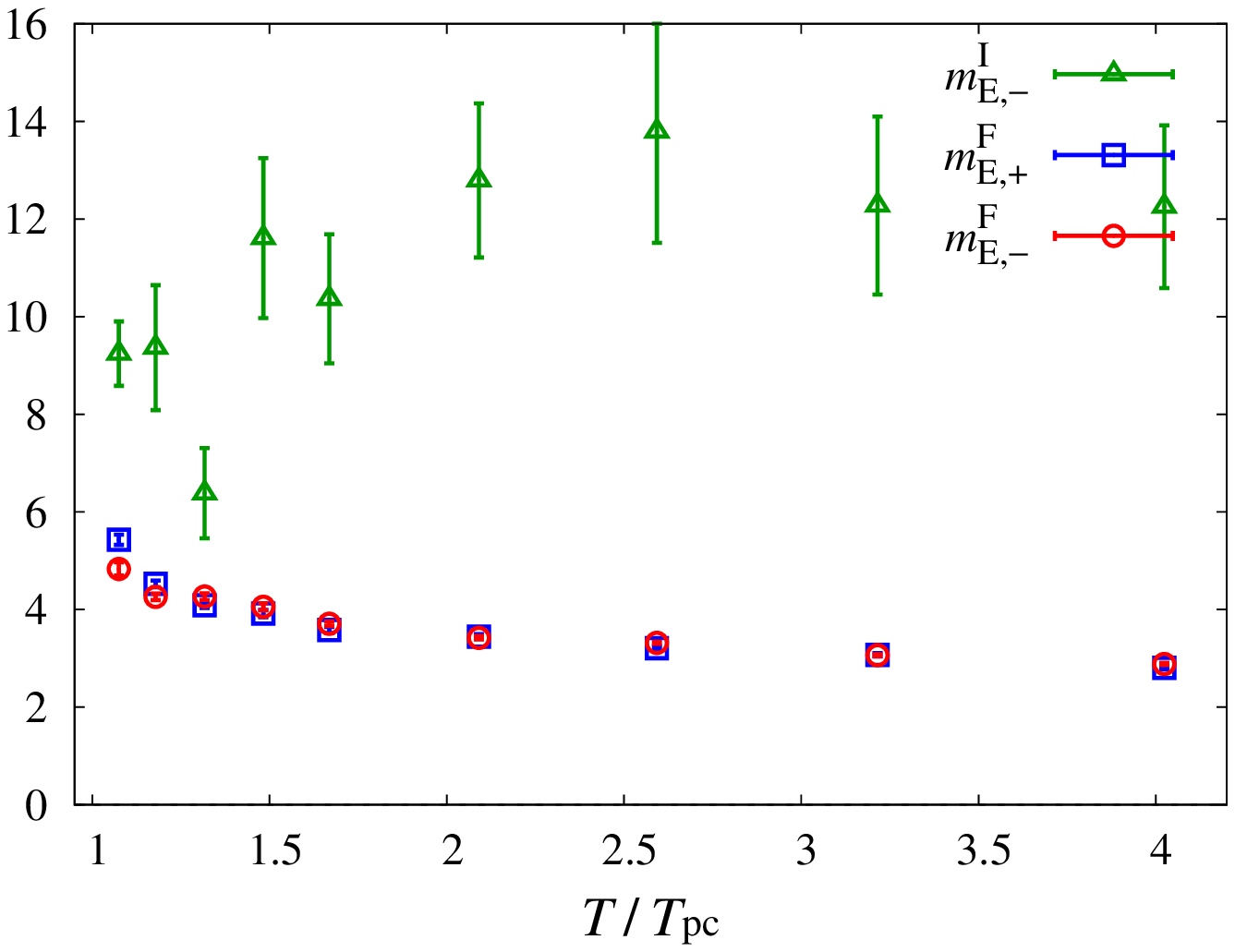}
    \end{tabular}
    \caption{Screening masses in magnetic (left)
    and electric (right) sectors at $m_{\rm PS}/m_{\rm V} = 0.65$.
        }
    \label{fig:1}
  \end{center}
  \vspace{-2mm}
\end{figure}

Left (right) panel of Fig.~\ref{fig:1} shows results of screening masses
 in magnetic (electric) sector at $m_{\rm PS} / m_{\rm V} = 0.65$
 as a function of temperature.
 As temperature increases,
  the magnetic  screening masses 
   converge to a similar value, whereas
  the electric screening masses shows 
   $\Meef \simeq \Meof < \Meoi$.
 In the weak coupling limit,
 these properties may be interpreted as follows.
The Polyakov-line operator in the magnetic (electric) sector
 has even (odd) powers of $g$:
\be
\Omega_{\rm M} &\sim& 1 + \f{1}{2} \left( i g \int_0^{1/T} d\tau A_4 \right)^2 + \cdots
,\\
\Omega_{\rm E} &\sim& - ig\int_0^{1/T} d\tau A_4 - \f{1}{3!} \left( i g \int_0^{1/T} d\tau A_4 \right)^3 + \cdots
.
\ee
Then all the Polyakov-line correlations in the magnetic sector start with the
 correlation of the second powers of $A_4$:
\be
C_{\rm M}^{\rm F} (r,T) &\propto& g^4 \la \tr A_4^2(\bx) A_4^2(\by) \ra + O(g^8)
,\\
C_{\rm M}^{\rm I} (r,T) &\propto& g^4 \la \tr A_4^2(\bx) \tr A_4^2(\by) \ra + O(g^8)
.
\label{eq:A2A2}
\ee
On the other hand, Polyakov-line correlations in the electric sector 
have different powers depending on the gauge choice,
because ${\rm Tr} A_4 =0$:
\be
C_{\rm E}^{\rm F} (r,T)  &\propto& g^2 \la \tr A_4(\bx) A_4(\by) \ra + O(g^6)
,\\
C_{\rm E}^{\rm I} (r,T)  &\propto& g^6 \la \tr A_4^3(\bx) \tr A_4^3(\by) \ra + O(g^8)
.
\ee
Then the leading contribution in $C_{\rm E}^{\rm F}$
 comes from single temporal-gluon exchange, whereas 
  that in $C_{\rm E}^{\rm I}$
 comes from triple temporal-gluon exchange.
This may lead to the exponential fall of  
$C_{{\rm E},-}^{\rm I} $ much  faster than 
$C_{{\rm E},\pm}^{\rm F}$ and hence 
 the mass ordering in Fig.\ref{fig:1}.
Further  discussions on a relation between the Polyakov-line correlations
 and the gluon propagators in the weak coupling are given in Appendix.

\subsection{Comparison with 3D effective field theory and ${\cal N}=4$
 supersymmetric Yang-Mills theory}

\begin{figure}[tbp]
  \vspace{-2mm}
  \begin{center}
    \includegraphics[width=80mm]{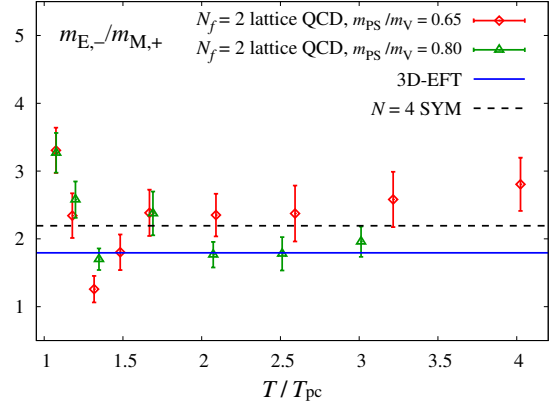} 
    \caption{Comparison of the screening ratio, $m_{{\rm E},-}/m_{{\rm M},+}$,
    with predictions in the dimensionally-reduced effective field theory (3D-EFT) \cite{Hart:2000ha}
    and ${\cal N}=4$ supersymmetric Yang-Mills theory (SYM) \cite{Bak:2007fk}. }
    \label{fig:2}
  \end{center}
  \vspace{-2mm}
\end{figure}

Let us compare our results with the predictions by
 the dimensionally reduced effective field theory (3D-EFT) 
  and the ${\cal N}=4$ supersymmetric Yang-Mills theory (SYM).
Here we focus our attention on the screening masses
 with $(\Te,C)=(+,+)$ and $(-,-)$,
 i.e. $m_{{\rm M},+}$ and $m_{{\rm E},-}$,
  which are channels calculable
  from the gauge invariant Polyakov-line correlations\footnote{
 Although our Polyakov-line operators do not have definite
  angular momentum $J$ and the parity $P$, 
  we assume that the asymptotic 
   behavior of  our Polyakov-line correlations picks up
    the ground state contribution corresponding to
     $(J,P)=(0,+)$  }.
In the 3D-EFT approach \cite{Hart:2000ha}, the screening masses 
at $T=2\Lambda_{\overline{MS}}$ with $\Lambda_{\overline{MS}}$ being
 the QCD scale parameter have  been calculated for $N_f=2$ as 
\be
\textrm{3D-EFT}\ (N_f=2):  \ m_{{\rm M},+} \sim 3.60T , \ \ m_{{\rm E},-} \sim 6.46T .
\ee
The screening masses of the Polyakov-line correlations
  in ${\cal N}=4$ supersymmetric Yang-Mills theory 
 in the limit of large $N_c$ and 
   large 'tHooft coupling ($\lambda = g^2 N_c$) were calculated
    by  AdS/CFT correspondence \cite{Bak:2007fk}:
\be
{\cal N}=4 \ \textrm{SYM}: \ m_{{\rm M},+} \sim 7.34T , \ \ m_{{\rm E},-} \sim 16.1T .
\ee
Our results of the screening masses  shown in Fig.~\ref{fig:1}
 indicates that
\be
\textrm{4D-lattice QCD} \ (N_f=2): \ m_{{\rm M},+} = 4T-6T , \ \ m_{{\rm E},-} = 10T-14T .
\ee
In all three cases, we have the inequality:
  $m_{{\rm M},+} < m_{{\rm E},-}$. Since we cannot 
  compare the absolute magnitude
   of the screening masses in QCD and those in
  ${\cal N}=4$ SYM because of the different number of degrees of freedom,
 we take the ratio, $m_{{\rm E},-}/m_{{\rm M},+}$ in Fig.~\ref{fig:2}
  for above three cases.
We find that, in spite of the difference in the magnitude, 
 the screening ratios agree well with each other
 for $1.5 < T/\Tpc < 3$.

\section{Summary}
\label{sec:summary}

We investigated the screening masses in hot quark-gluon plasma
 from the Polyakov-line correlations classified by
 the Euclidean-time reflection ($\Te$) symmetry 
  and the charge conjugation ($\C$) symmetry.
 The screening masses are calculated 
 in the magnetic ($\Te$-even) and electric ($\Te$-odd) sectors
   from the lattice simulations of $N_f=2$ QCD above $\Tpc$
 with the RG-improved gluon action and
  the clover-improved Wilson quark action
   on a $16^3 \times 4$ lattice.
We found that,
  the magnetic screening masses obtained 
   by the gauge-invariant and gauge-fixed correlations coincide
    with each other for $T/\Tpc >2$, whereas
   the electric screening mass obtained from the gauge invariant correlator
    is larger than that obtained from the gauge fixed correlator.
 Our results obtained from the gauge invariant correlators
  are compared with the calculations in
   the dimensionally-reduced effective field theory \cite{Hart:2000ha} and 
 the ${\cal N}=4$ supersymmetric Yang-Mills theory \cite{Bak:2007fk}.
  We found that the ratio of the electric and magnetic  screening masses 
 in three cases are qualitatively consistent with each other for
    at $1.3 < T/\Tpc < 3$.
 Further discussions  between the Polyakov-line correlations
 and the gluon propagators are given in Appendix.

\paragraph{Acknowledgements:}
We would like to thank M.~Laine for giving useful advice
 about the Euclidean time symmetry.
This work is in part supported by Grants-in-Aid of the Japanese Ministry 
of Education, Culture,
Sports, Science and Technology (Nos. 17340066, 18540253, 19549001, and 20340047). 
SE is supported
by U.S. Department of Energy (DE-AC02-98CH10886).
Numerical calculations were performed on supercomputers at KEK by the
Large Scale Simulation Program Nos.\ 06-19, 07-18, 08-10, at CCS,
Univ.\ of Tsukuba, and at ACCC, Univ.\ of Tsukuba.

\appendix
\section{Appendix: Relation to gluon propagators}

In this Appendix, we consider a direct relation between 
 the Polyakov-line correlations and the screening mass in gluon propagators.
 Here we define the magnetic (electric) 
 screening mass, $m_M$ $(m_E)$, from the spatial (temporal) gluon propagator
 at  asymptotic spatial distance,
\be
\la A_i (\bx) A_i (\by) \ra &\rightarrow& T \  \frac{e^{-m_M r}}{4 \pi r} 
 \ \ \   (|\bx - \by| = r \rightarrow \infty), 
 \\
\la A_4 (\bx) A_4 (\by) \ra &\rightarrow& T \ \frac{e^{-m_E r}}{4 \pi r}
 \ \ \   (|\bx - \by| = r \rightarrow \infty).
\ee
In the weak coupling limit, the gauge fixed correlation
 $\Ceof$ is related to the electric propagator as 
\be
\Ceof (r,T) \rightarrow a(T) \f{e^{-m_E(T) r}}{rT}
.
\label{eq:Ceof}
\ee
On the other hand, the gauge invariant correlation $\Cmei$ 
 would have not only an exchange of two temporal gluons
   but also an exchange of  two spatial gluons \cite{Arnold:1995bh}:
\be
\Cmei (r,T) \rightarrow b_1(T) \left( \f{e^{-m_E(T) r}}{rT} \right)^2 
            + b_2(T) \left( \f{e^{-m_M(T) r}}{rT} \right)^2  .
\ee
Although $b_2$ has higher powers in $g$, the exchange of the 
 spatial gluons could dominate at long distances if $m_M < m_E$. 

We extract $m_E$ by fitting $\Ceof$
 to the right-hand-side of Eq.~(\ref{eq:Ceof}) 
  with the fitting parameters, $a(T)$ and $m_E(T)$.
Then we extract $m_M$ by fitting a combination 
 of the $\Ceof$ and $\Cmei$ as,
\be
\f{\Cmei(r,T)}{(\Ceof(r,T))^2} 
 = c_1 + c_2(T) \exp \left[ 2 \left( m_E(T) - m_M(T) \right) r \right]
,
\label{eq:Ce_o}
\ee
with $c_1$, $c_2$ and $m_M$ being the fitting parameters.
The fit ranges are chosen to be $0.5 \le rT \le 1.25$.

\begin{figure}[tbp]
  \vspace{-2mm}
  \begin{center}
    \begin{tabular}{cc}
    \includegraphics[width=74mm]{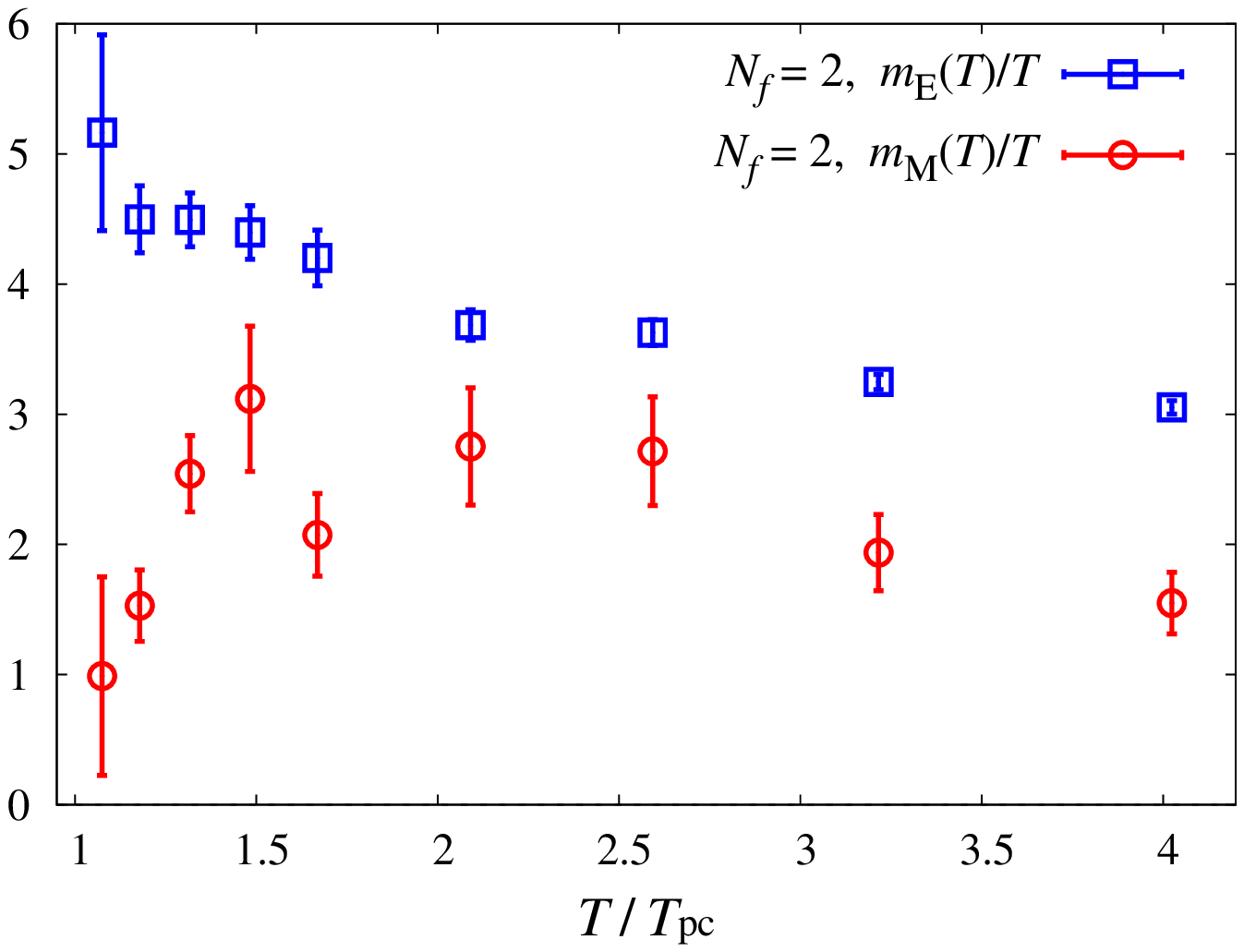} &
    \includegraphics[width=74mm]{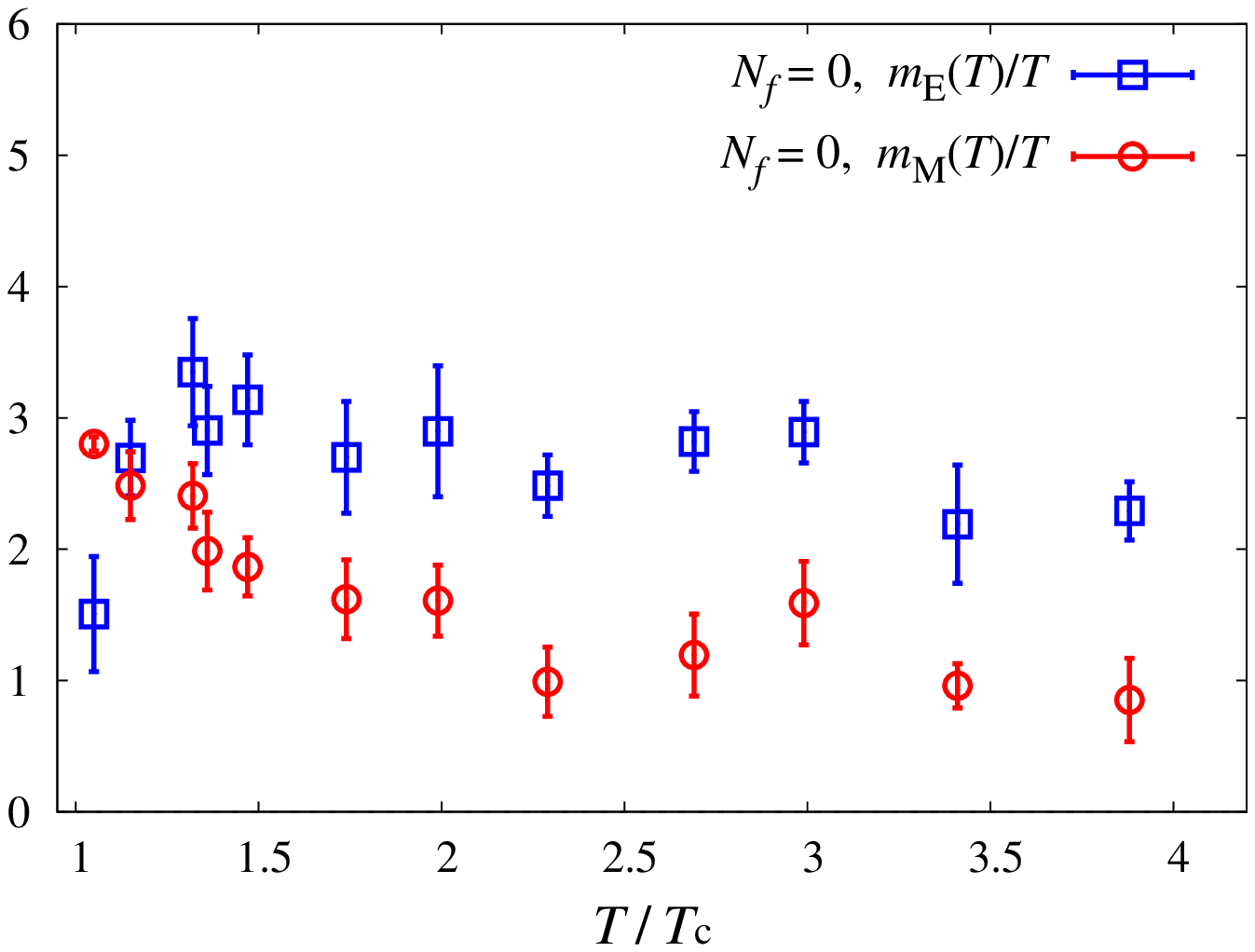}
    \end{tabular}
    \caption{Results of the electric and magnetic screening masses
     calculated from the Polyakov-line correlations 
      in $N_f=2$ lattice QCD at $m_{\rm PS}/m_{\rm V} = 0.65$ (left)
      and from the 
      gluon propagators in $N_f=0$ lattice QCD  \cite{Nakamura:2003pu} (right).
        }
    \label{fig:3}
  \end{center}
  \vspace{-2mm}
\end{figure}

Results of $m_E(T)$ and $m_M(T)$ 
 are shown in Fig.~\ref{fig:3} (left)
 as a function of $T/\Tpc$ at $m_{\rm PS} / m_{\rm V} = 0.65$.
Similar results are obtained for $m_{\rm PS} / m_{\rm V} = 0.80$ too.
We find that  $m_M$ is smaller
  than $m_E$ at all temperatures we calculate,
   which is consistent with a prediction in 
   the thermal perturbation theory:
    $m_M \sim O(g^2T) < m_E \sim O(gT)$.
We also find that, for $T/\Tpc > 2 $, both screening masses 
 decreases as $T$ increases, whereas $m_E$ and $m_M$
 behaves differently for $1 < T/\Tpc < 2$, 
  i.e. $m_E$ increases and $m_M$ decreases
    when the temperature approaches to $\Tpc$.
The behavior of $m_M$ is  not expected 
 from the leading-order perturbation theory.

 In the right panel of Fig.~\ref{fig:3},
 $m_E$ and $m_M$  calculated directly from the gluon propagators
  in   the quenched approximation are recapitulated \cite{Nakamura:2003pu}.
 Although  $m_M < m_E$ is also found for $T/T_{\rm c} \simge 1.2$,
  $m_M$ ($m_E$)increases (decreases)
   as $T$ approaches to $T_{\rm c}$ in the quenched calculation.
 Such qualitative difference between $N_f =2$  and $N_f =0$
  near the (pseudo-) critical temperature
 may be attributed to the different behavior of the 
 the phase transition in two cases.

\end{document}